\documentclass[%
 reprint,
superscriptaddress,
 amsmath,amssymb,
 aps,
]{revtex4-2}

\usepackage{graphicx}
\usepackage{dcolumn}
\usepackage{bm}
\usepackage{tabularx}
\usepackage{booktabs}
\usepackage{multirow}
\usepackage{makecell}  
\usepackage{threeparttable}

\begin{document}

\preprint{APS/123-QED}

\title{Photonic-Implemented Efficient Deep Quantum Neural Network via Virtual-Driven Hilbert Space Expansion}

\author{Haoran Ma}
\thanks{These authors contributed equally to this work.}
\affiliation{
	College of Information Science and Electronic Engineering and
	ZJU-Hangzhou Global Scientific and Technological Innovation Center,
	Zhejiang University, Hangzhou, Zhejiang, China
}

\author{Huihui Zhu$^{*}$}
\email{huihui001@zju.edu.cn}
\affiliation{
	College of Information Science and Electronic Engineering and
	ZJU-Hangzhou Global Scientific and Technological Innovation Center,
	Zhejiang University, Hangzhou, Zhejiang, China
}

\author{Zichao Zhao}
\affiliation{
	College of Information Science and Electronic Engineering and
	ZJU-Hangzhou Global Scientific and Technological Innovation Center,
	Zhejiang University, Hangzhou, Zhejiang, China
}

\author{Qishen Liang}
\affiliation{
	College of Information Science and Electronic Engineering and
	ZJU-Hangzhou Global Scientific and Technological Innovation Center,
	Zhejiang University, Hangzhou, Zhejiang, China
}

\author{Liao Ye}
\affiliation{
	College of Information Science and Electronic Engineering and
	ZJU-Hangzhou Global Scientific and Technological Innovation Center,
	Zhejiang University, Hangzhou, Zhejiang, China
}

\author{Baojie Hou}
\affiliation{
	College of Information Science and Electronic Engineering and
	ZJU-Hangzhou Global Scientific and Technological Innovation Center,
	Zhejiang University, Hangzhou, Zhejiang, China
}

\author{Jia Guo}
\affiliation{
	College of Information Science and Electronic Engineering and
	ZJU-Hangzhou Global Scientific and Technological Innovation Center,
	Zhejiang University, Hangzhou, Zhejiang, China
}

\author{Leong Chuan Kwek}
\email{cqtklc@nus.edu.sg}
\affiliation{
	Centre for Quantum Technologies,
	National University of Singapore,
	Singapore, Singapore
}

\author{Mile Gu}
\affiliation{
	Nanyang Quantum Hub,
	School of Physical and Mathematical Sciences,
	Nanyang Technological University,
	Singapore 639798, Singapore
}

\author{Jayne Thompson}
\affiliation{
	College of Computing and Data Science,
	Nanyang Technological University,
	Singapore 639798, Singapore
}

\author{Wei Luo}
\affiliation{
	Research Institute for Quantum Technology (RIQT),
	The Hong Kong Polytechnic University,
	Hung Hom, Hong Kong, China
}

\author{Yuehai Wang}
\affiliation{
	College of Information Science and Electronic Engineering and
	ZJU-Hangzhou Global Scientific and Technological Innovation Center,
	Zhejiang University, Hangzhou, Zhejiang, China
}

\author{Jianyi Yang}
\email{yangjy@zju.edu.cn}
\affiliation{
	College of Information Science and Electronic Engineering and
	ZJU-Hangzhou Global Scientific and Technological Innovation Center,
	Zhejiang University, Hangzhou, Zhejiang, China
}
\affiliation{
	Jinhua Institute of Zhejiang University,
	Jinhua 321002, Zhejiang, China
}

\begin{abstract}
The growing computational demands of classical neural networks have intensified the search for energy-efficient and powerful computational alternatives. Quantum neural networks (QNNs) implemented on integrated photonic platforms offer a compelling avenue, offering exceptional computational power enhancements, with inherent programmability and scalability of integrated architectures. A critical challenge, however, is implementing the fundamental non-unitary and nonlinear activation function of QNNs within a linear quantum photonic system. Existing strategies, such as the adding ancillary qubits and measurement-based feedback or forward are constrained by high qubit resource costs, overhead devices, and poor cascadability. Here, we propose a novel deep photonic QNN with an expanded computational Hilbert space via input replication and mode expansion, which enables the realization of effective non-unitary and nonlinear activation on a linear programmable quantum photonic chip. This approach eliminates the need for physical ancillary qubits, measurement-induced qubit consumption and the measurement device burden, thereby significantly reduce resource costs. The fabricated chip integrates four high-quality entanglement sources and a programmable high-dimensional interferometric network, enabling a two-hidden-layer QNN that exhibits dimension-enhanced expressivity over the existing QNN architectures. We demonstrate its capabilities across diverse tasks, including nonlinear classification, image generation, and quantum Gibbs state preparation. This work establishes a scalable and efficient architecture toward practical quantum deep learning systems capable of tackling problems beyond the reach of classical computation.
\end{abstract}

\maketitle

The explosive growth in neural network size has dramatically increased energy consumption and training costs, spurring the search for more efficient alternatives. This has driven the development of novel hardware platforms, including analog electronics \cite{indiveri2011neuromorphic}, optical computing \cite{zhu2022space}, and quantum neural networks (QNNs) \cite{abbas2021power}. Among these, QNNs stand out by leveraging the exponential scaling of the Hilbert space with the number of qubits, combining quantum superposition with parallel neural processing to potentially revolutionize machine learning \cite{bharti2022noisy}. Recently, there have been efforts toward QNN proposals, such as quantum kernel methods \cite{jerbi2023quantum}, data replication (e.g. reuploading) circuits \cite{perez2020data, schuld2021effect}, and shadow models \cite{jerbi2024shadows}, sparking the rapidly developing field of QNNs, based on parameterized quantum circuits in the era of noisy intermediate-scale quantum (NISQ) computers \cite{benedetti2019parameterized,biamonte2017quantum}. 

QNNs, similar to classical neural networks, connect neurons through linear maps and nonlinear activation functions. So far, physical systems have been capable of implementing linear unitary maps through the dynamics of the exponentially large many-qubit Hilbert space in superconducting circuits \cite{haug2023quantum,peters2021machine,havlivcek2019supervised}, trapped ions \cite{zhu2019training}, and photonic systems \cite{arrazola2021quantum}. Among the many QNN platforms, integrated photonic chips, as an important platform with strong integration and flexible programmability, combining the inherent speed of light propagation, low decoherence, and room-temperature operation, underscore their potential as a promising platform. Their high integration and flexible programmability enable many linear quantum operations, including the generation of high-dimensional entangled states \cite{wang2018multidimensional,llewellyn2020chip}, graph states \cite{bao2023very, zhu2024dynamically, huang2024demonstration}, and Gaussian boson sampling \cite{yu2023universal,huh2015boson}. These capabilities firmly establish integrated photonics as a compelling approach for implementing QNNs \cite{hoch2025quantum,shang2024codesigned,xue2022variational,carolan2020variational}.

However, implementing QNNs on integrated quantum photonic platforms remains a fundamental challenge, due to the weak interaction between photons \cite{knill2001scheme,bartolucci2023fusion}. This physical limitation creates two primary bottlenecks: the linear unitary gates are typically implemented probabilistically, and robust nonlinear activation functions remain difficult to realize. While the KLM scheme \cite{knill2001scheme} and fault-tolerant protocols \cite{bartolucci2023fusion} provide theoretical paths toward universal quantum computing with probabilistic gates under linear optics, they incur prohibitive experimental overhead, such as stringent photon loss budgets and high multiplexing depths\cite{psiquantum2025manufacturable}. To bypass the lack of native nonlinearity, current techniques often synthesize nonlinear behavior through ancillary qubits and measurement-induced projection\cite{wan2017quantum}. However, the resulting expansion of the Hilbert space demands significant computational resources,  rendering these approaches impractical for large-scale applications or scenarios requiring strong nonlinearities \cite{jerbi2023quantum}. Alternatively, measurement-based feedback schemes have been theoretically proposed \cite{hoch2025quantum,knill2001scheme,magann2022feedback}. These require real-time feedback control of quantum operations based on measurement outcomes, a process that necessitates challenging active modulation. Although recent research uses post-selection to reduce this experimental burden \cite{hoch2025quantum}, the inherent loss of qubits through measurement or post-selection imposes severe layer-depth constraints. Consequently, both ancillary qubit and feedback-based methods place a substantial burden on qubit resources, limiting their practicality for demonstrating deep QNNs. Therefore, a key requirement for realizing practical QNNs is to improve probability of linear operation and minimize both the consumption of physical qubits and the need for real-time feedback control. Photons inherently meet this need by offering multiple degrees of freedom, which provide a powerful pathway for scaling quantum computations. The synthetic dimension approach, for instance, facilitates the construction of additional Hilbert spaces beyond the defacto computational subspace to enable high-dimensional quantum simulations \cite{liu2025reconfigurable, monika2025quantum}. Leveraging this expanded dimensional space reduces the reliance on unitary gates and physical ancillary qubits, thereby facilitating the implementation of efficient, deep QNNs.

In this paper, we propose a nontrivial protocol and demonstrate a highly entangled programmable silicon-photonic quantum chip for the ancillary dimensional expansion-assisted quantum neural network (ADE-QNN). Our new architecture uniquely eliminates the need for introducing additional physical systems to act as acillary qubits or feedback loops by leveraging input state replication and mode extension to generate a high-dimensional, virtual computation space. Non-unitary and nonlinearity are efficiently induced through measurements on this ancillary space, a scheme that improves probability of linear operation and consumes no additional physical qubit or diminishes the original computational qubits. This approach enables a cascaded framework for construcing deep QNNs that is much more scalable and resource-efficient. The fabricated chip comprises entanglement resources and a high-dimensional interference network with high fidelity and exceptional programmability to experimentally implement the two-hidden-layer ADE-QNN. This device successfully performs classification of highly nonlinear datasets and, more notably, unlocks complex tasks such as the generation of MNIST images and the learning of quantum Gibbs states. Compared with previous QNNs, our demonstration of a virtual-dimensions-assisted architecture facilitates an unprecedented, deep QNN without proportional scaling of physical qubits. Our scheme and integrated quantum photonic chip pave the way for large-scale, multi-layered QNNs, marking a significant stride toward practical quantum machine learning applications for next-generation systems.

\begin{figure*}[t]
	\centering
	\includegraphics[width=1\textwidth]{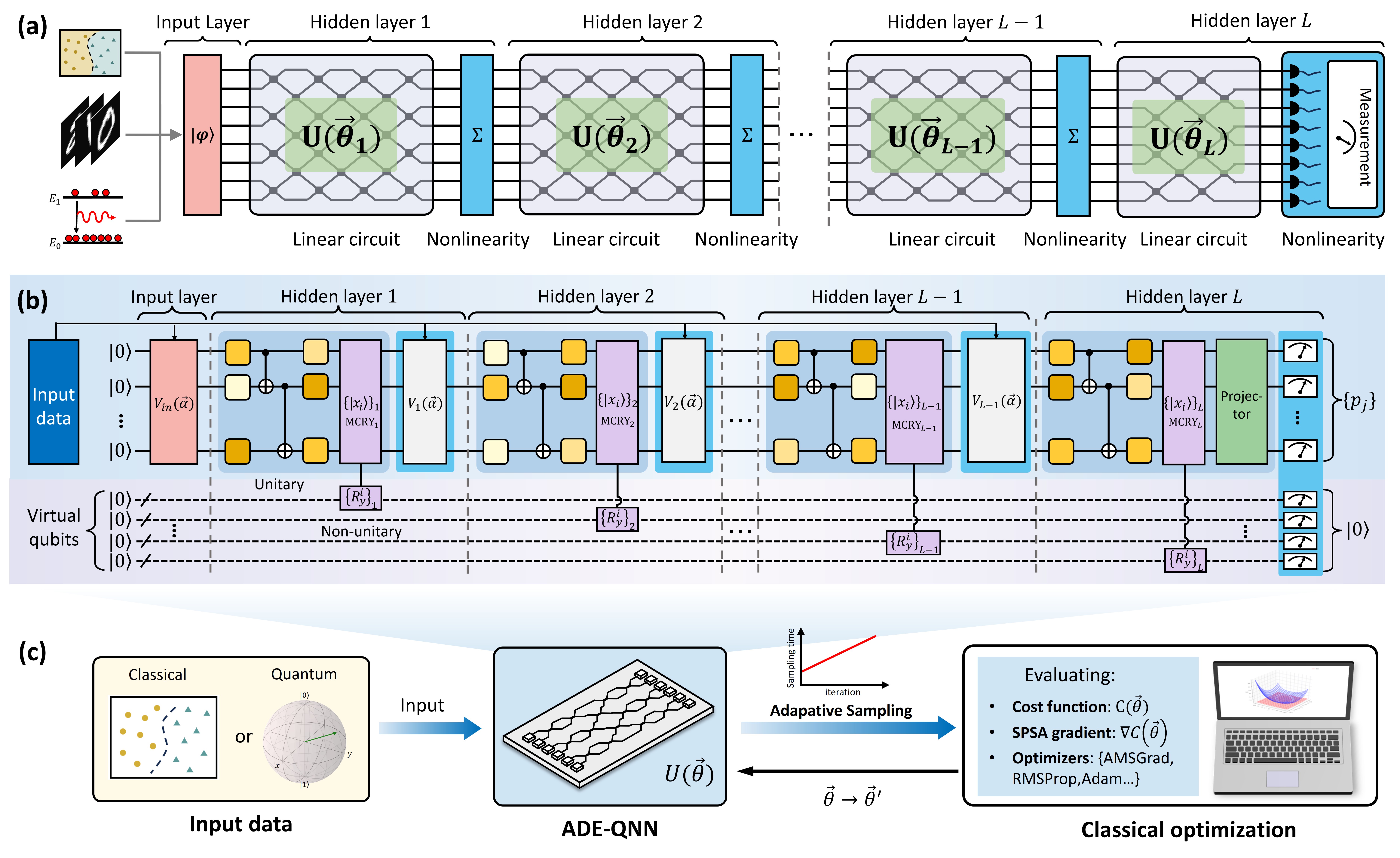}
	\caption{\textbf{Architecture of the ADE-QNN.} (a) General framework of the ADE-QNN. It includes an input layer, $L$ hidden layers, and a final single-photon measurement. The target data (classical decision boundary datasets, MNIST images, and quantum thermoequilibrium Gibbs states are shown here) are loaded into the ADE-QNN in the form of a quantum photonic state $|\varphi\rangle$, which then undergoes nonlinear evolution in the $L$ hidden layers parameterized with $\vec{\theta}$. The results are read out in the final single-photon measurement. (b) The detailed quantum circuit implementing the ADE-QNN. The input layer is a quantum circuit $V_{in}(\vec{\alpha})$, where $\vec{\alpha}$ is the data vector. The non-unitary part of linear circuit is performed by an MCRY operation, which introduces an additional virtual qubit set. The nonlinear activation is implemented by replicating the input data with $V_k(\vec{\alpha})$ and measurement. The quantum circuit is partially measured by post-selecting the virtual qubits in the state $|0\rangle$. The computational qubits (blue background) serve as readout qubits, presenting measurement results $\{p_j\}$ with $\{|j\rangle\}$ being the set of basis states. (c) The ADE-QNN optimization workflow. The ADE-QNN receives input data and reads out the computational results using an adaptive sampling strategy. These results are sent to a classical computer to evaluate a problem-specific cost function $C(\vec{\theta})$ and its gradients $\nabla C(\vec{\theta})$. The gradients are used to update the parameters $\vec{\theta}$ of the ADE-QNN via an optimizer. This hybrid quantum-classical optimization loop iterates until a convergence criterion is met.
	}\label{fig1}
\end{figure*}

\vspace{1em}
\noindent \textbf{Architecture of the ADE-QNN}

The overarching architecture of the ADE-QNN with multi-layers is depicted in Figure 1(a), which is composed of an input layer, hidden layers with each consisting of a linear and a nonlinear layer, and single-photon measurements with feedback-driven parameter optimization to refine the learning process. In this work, diverse classical and quantum datasets, including two-dimensional Circle and Spiral, Glass patterns with three labels, MNIST images, and quantum Gibbs states, are first uploaded into the circuit in the form of quantum state $|\varphi \rangle$. Specifically, classical data is mapped into the angles of rotation gates, which encodes the data into quantum states. The linear part of hidden layers is implemented via a parameterized quantum circuit, $U(\vec{\theta})$, constructed from multiple tunable single-qubit and two-qubit gates and the dimension expansion. The parameters $\vec{\theta}$, which define the circuit's operation within the $n$-qubit Hilbert space, are optimized through training. The nonlinear part of the inside hidden layers $\Sigma$ is implemented by replicating the input state. Finally, a photon measurement is performed on the readout qubit in the final hidden layer, introducing futher nonlinearity. This ADE-QNN can be physically implemented on a linear-optics-based platform without relying on ancillary photons, mid-circuit measurements, and fast feedback, thereby greatly enhancing experimental feasibility for quantum machine learning.

Figure 1(b) depicts the detailed quantum circuit implementing the ADE-QNN scheme. It includes $L+1$ layers of quantum circuits (1 input layer and $L$ hidden layers). Each hidden layer has two subsections, i.e., the linear quantum circuit and the nonlinear activation function. The unitary parts of the linear subsections are equipped with parameterized gates and entangling operations between qubits in the computational register, the parameters of which need to be determined through training. The non-unitary operation of the linear subsections is achieved through a dimension-extension-based, multivalue-controlled RY (MCRY) module, where RY indicates the rotation single-qubit gate along the y axis. Detailly, the dimension-extension-based MCRY module expands the Hilbert space, and a subsequent partial measurement on this expanded space introduces additional non-unitary. This process is equivalent to applying a collection of operations $\{R_y^i\}_k$ on
		the $k$-th set of virtual qubits, with each $R_y^i$ controlled by basis state $|x_i\rangle$ in the computational register and the number of virtual qubits per set determined by the size of the expanded space. The detailed analysis about non-unitary expressivity arising from dimension expansion is in Supplementary Note 1. The nonlinearity is provided via an encoding module ($V(\vec{\alpha}_k)$) that replicates the input data state. Among them, it is possible to show that this replication is functionally equivalent to expanding the $n$-qubit input state to a $(n+n_a)$-qubit state ($n_a\geq n$) followed by a partial measurement on $n$ qubits, which induces the necessary nonlinearity (see details in Supplementary Note 1). Therefore, the full system comprising $L$ layers is
\begin{equation}
		S = U(\vec{\theta}_{L})\left(\prod_{k=1}^{L-1}V_k(\vec{\alpha})U(\vec{\theta}_k)\right)V_{in}(\vec{\alpha}),
\end{equation}

\noindent where $U(\vec{\theta}_k)$ and $V_k(\vec{\alpha})$ represent the linear transformation and the effect of the encoding module within each layer, respectively. Finally, a projector module is placed after the $L$ layers for quantum state tomography, following a quantum measurement $\hat{M}$, where the output state is post-selected on the virtual qubits being measured in the state $|0\rangle$, enabling partial state measurement. The measurement of the quantum circuits leads to
\begin{equation}
	\{p_j\} = \left\{Tr\left[|j\rangle\langle j|Tr_v\left(\frac{\Pi S|0\rangle\langle0|S^{\dagger}\Pi}{Tr(\Pi S|0\rangle\langle0|S^{\dagger})}\right)\right]\right\},
\end{equation}

\begin{figure*}[!t]
	\centering
	\includegraphics[width=1\textwidth]{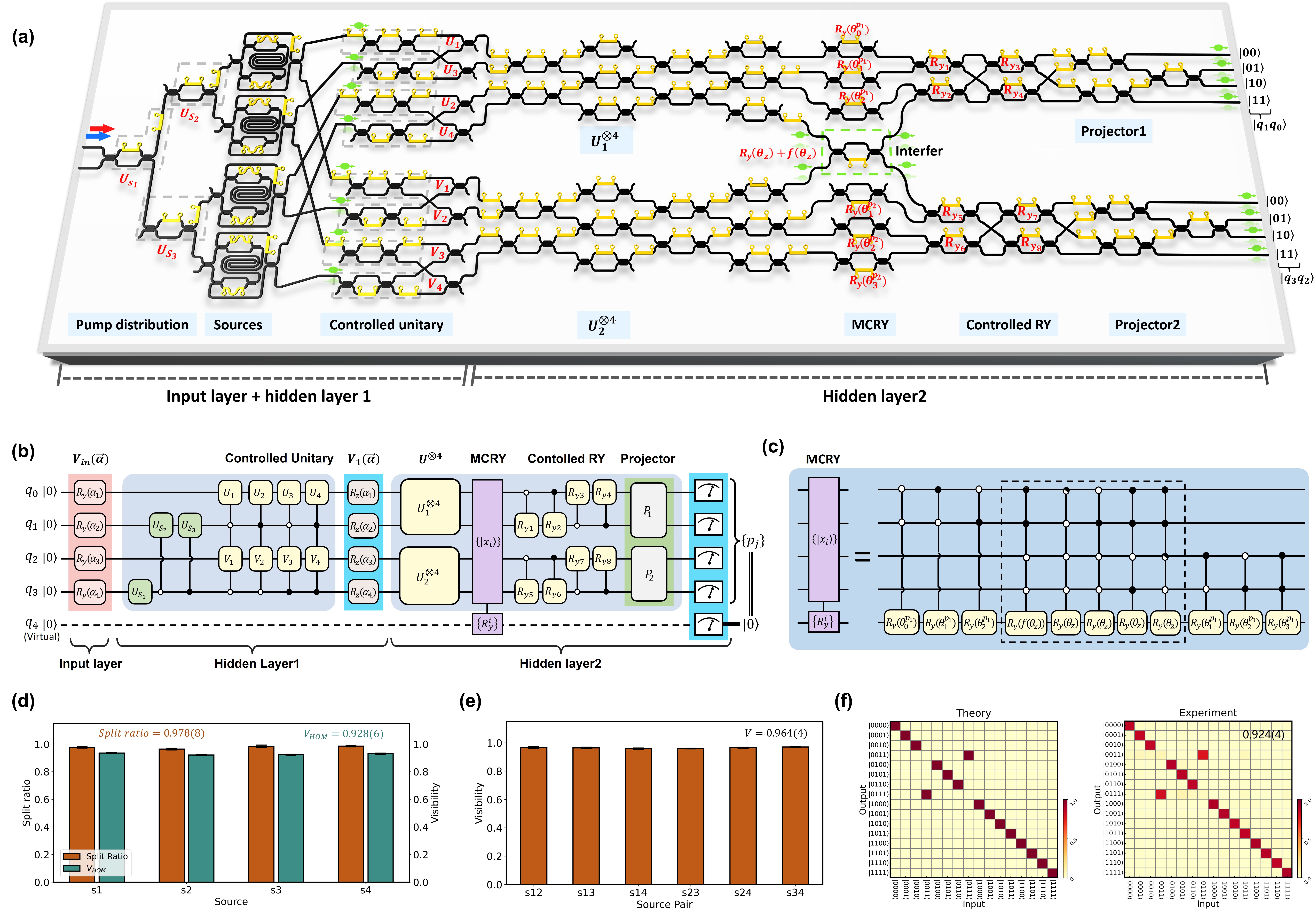}
	\caption{\textbf{A quantum photonic chip for realizing ADE-QNN.} (a) Schematic of the designed quantum photonic chip and (b) the equivalent quantum circuit. It includes two hidden layers, where the first and second nonlinear layers are realized with data replication and the measurement, respectively. The input layer and the 1st hidden layer are implemented with a combination of pump distribution, sources, and controlled unitaries. The 2nd hidden layer is performed using the $U^{\otimes4}$, MCRY operations, controlled RY gates, two projectors and single-photon measurements. A dual pump (red and blue arrows) is used to stimulate photon pairs (green arrows), where photon 1 and photon 2 serve as $q_1q_0$ and $q_3q_2$, respectively. The green box with a dotted line indicates where two-photon interference may occur. (c) Detailed description of the MCRY module. $\theta_z$ has two distinct values in our ADE-QNN experiments, and $f(\theta_z)$ is a piecewise function of $\theta_z$. (d) Brown bars: measured split ratio for the four sources, with a mean of 0.978(8); green bars: measured photon indistinguishability via on-chip Hong-Ou-Mandel (HOM) experiment on the MZI with $\theta_z$, showing a mean visibility of 0.928(6). (e) Measured indistinguishabilities among different combinations of source pairs, having a mean visibility of 0.964(4). (f) The theoretical and experimental truth table for the CCCX gate in the MCRY module, showing a statistical fidelity of 0.924(4).
	}\label{fig2}
\end{figure*}

\noindent where $\Pi=(|0\rangle\langle 0|)^{\otimes v} \otimes \mathcal{I}$ defines the measurement projector, with $\mathcal{I}$ the identity operator, $\{|j\rangle\}$ represent the set of basis states of the computational qubits, and $\{p_j\}$ are the measured probabilities. To efficiently train this ADE-QNN, the cost function is defined as $C(\vec{\theta})$, which can be efficiently calculated from $\{p_j\}$. The goal of the training is to minimize $C(\vec{\theta})$, which is performed using a feedback training loop shown in Figure 1(c). The measurement results are sent to a classical computer to estimate both the cost function $C(\vec{\theta})$ and its gradients. To reduce physical training time, three approaches are employed: 1) adaptive sampling, where the sampling time for estimating $\{p_j\}$ increases with training iterations (shown in the inset of Figure 1(c)); 2) approximate estimation of the gradient $(\nabla C(\theta))$ using Simultaneous Perturbation Stochastic Approximation (SPSA); and 3) optimizer-combined parameter updating, i.e., AMSGrad and RMSProp \cite{wiedmann2023empirical}. A detailed description of these three techniques is provided in Supplementary Note 2. Using these strategies, the time required for gradient acquisition is greatly reduced while maintaining high precision, allowing for efficient training. Ultimately, by combining classical optimization with our parameterized quantum scheme, which fully implements classical/quantum data encoding, the linear network, and nonlinear activation function, we efficiently train the ADE-QNN and evaluate its performance on various tasks.

\vspace{1em}
\noindent \textbf{Design and validation of the quantum photonic chip} 

To achieve the aforementioned ADE-QNN, we designed an integrated quantum photonic chip, as illustrated in Figure 2(a). It is composed of integrated quantum sources and linear optical components, implementing an ADE-QNN with two hidden layers. The device consists of three main parts, namely: a multipartite entanglement operation (for the input layer and 1st hidden layer) and a combination of qudit unitary operations, an MCRY operation, a final combined controlled RY operation and projection measurement (for 2nd hidden layer). The equivalent quantum circuits are shown in Figure 2(b) and 2(c). A four-dimensional entanglement state with adjustable coefficients, $|\psi_0\rangle = \alpha|0000\rangle+\beta|0110\rangle+\gamma|1001\rangle+\delta|1111\rangle$, is first generated via integrated degenerate four-wave mixing sources with bidirectional pumping \cite{ma2025scheme}. Here $\alpha$, $\beta$, $\gamma$, and $\delta$ are tunable complex coefficients controlled by the pump power distribution module. In this configuration, this four-qubit state is mapped onto a two-photon system: the first photon's ququart state represents the qubit pair $q_1q_0$, while the second photon represents $q_3q_2$. Further, for the ququart modes, the Hilbert space expansion and coherent compression processes \cite{paesani2017experimental, qiang2018large, chi2022programmable, huang2024demonstration, ma2024quantum} are applied to transform $|\psi_0\rangle$ into the multipartite entanglement state as,

\begin{align}
	|\psi_1\rangle &= \alpha|00\rangle_{31} \otimes (V_1\otimes U_1)|00\rangle_{20} \nonumber \\
	&\quad + \beta|01\rangle_{31} \otimes (V_2\otimes U_2)|00\rangle_{20} \nonumber \\
	&\quad + \gamma|10\rangle_{31} \otimes (V_3\otimes U_3)|00\rangle_{20} \nonumber \\
	&\quad + \delta|11\rangle_{31} \otimes (V_4\otimes U_4)|00\rangle_{20},
\end{align}

\noindent where $U_i$ and $V_i$ are unitary operations applied on the target qubit pair $\{0,2\}$, controlled by the qubit pair $\{1,3\}$. This finished multipartite entanglement module achieves input layer ($V_{in}(\vec{\alpha})$), the 1st linear layer and a nonlinear activation function via input data replication ($V_{1}(\vec{\alpha})$), as detailed in the corresponding section of Figure 2 (b). The $V_{in}(\vec{\alpha})$ and $V_{1}(\vec{\alpha})$ modules are realized with rotation single-qubit gates $R_y$ and $R_z$, respectively, which are directly extracted from the chip's multipartite entanglement module, requiring no additional hardware overhead. It is noted that the combination of tunable entanglement sources and local unitary operations forms an efficient, highly entangled module, providing a large parameter space for the ADE-QNN's input layer and the 1st hidden layer. A detailed analysis of the tunable quantum source, entanglement state generation, controlled unitary operation and the input data encoding method is provided in Supplementary Note 3. The linear unitary part of the 2nd hidden layer is implemented by performing two arbitrary $U^{\otimes4}$ unitary operations, physically realized by universal linear-optical circuits \cite{clements2016optimal,zhao2024clements}, and a controlled RY operation. Besides, to further introduce non-unitary expressivity, an MCRY scheme is employed to be equal to two virtual qubits ($q_{v_1}$ and $q_{v_2}$, collectively abbreviated as $q_{\bm{v}}$). This is implemented by applying a unitary operation acting between path '11' of photon 1 and path '00' of photon 2, while the remaining paths are expanded for additional unitary operations. Through this process, the Hilbert space is expanded to four times its original dimensionality by increasing the path degrees of freedom of the photons. The subsequent tracing out of the ancillary modes projects the state into a lower-dimensional subspace, inducing a non-unitary transformation on the remaining system in the high-dimensional space. This scheme, equivalent to the MCRY part of the quantum circuit in Figure 2(b) and detailed in Figure 2(c), implements RY gates on $q_{\bm{v}}$ controlled by $q_0$–$q_3$, with subsequent post-selection on $|0\rangle$. This allows for multi-qubit controlled  unitary gates and a non-unitary operation enabled by the virtual qubit’s partial measurement. This MCRY, combined with two arbitrary local unitary operations $U^{\otimes4}$ (see Figure 2(a)) and a controlled RY operation, forms a module that provides a globally entangled linear transformation and non-unitary for the 2nd hidden layer. Details of the state evolution and experimental partial measurement are provided in Methods. Finally, a projective measurement provides nonlinearity for the 2nd hidden layer. 

The $1.8\times 8 mm^2$-sized quantum photonic chip is fabricated on silicon using the complementary metal oxide-semiconductor technology, and monolithically integrates 4 photon pair sources, 109 thermo-optic phase shifters, 136 multi-mode interferometer beam splitters, and 68 optical couplers. It is pumped by two continuous-wave lasers to generate frequency-degenerate photon pairs. Details for chip image, device fabrication, and experimental setup have been provided in Methods and Supplementary Note 4.

\begin{figure*}[!t]
	\centering
	\includegraphics[width=1\textwidth]{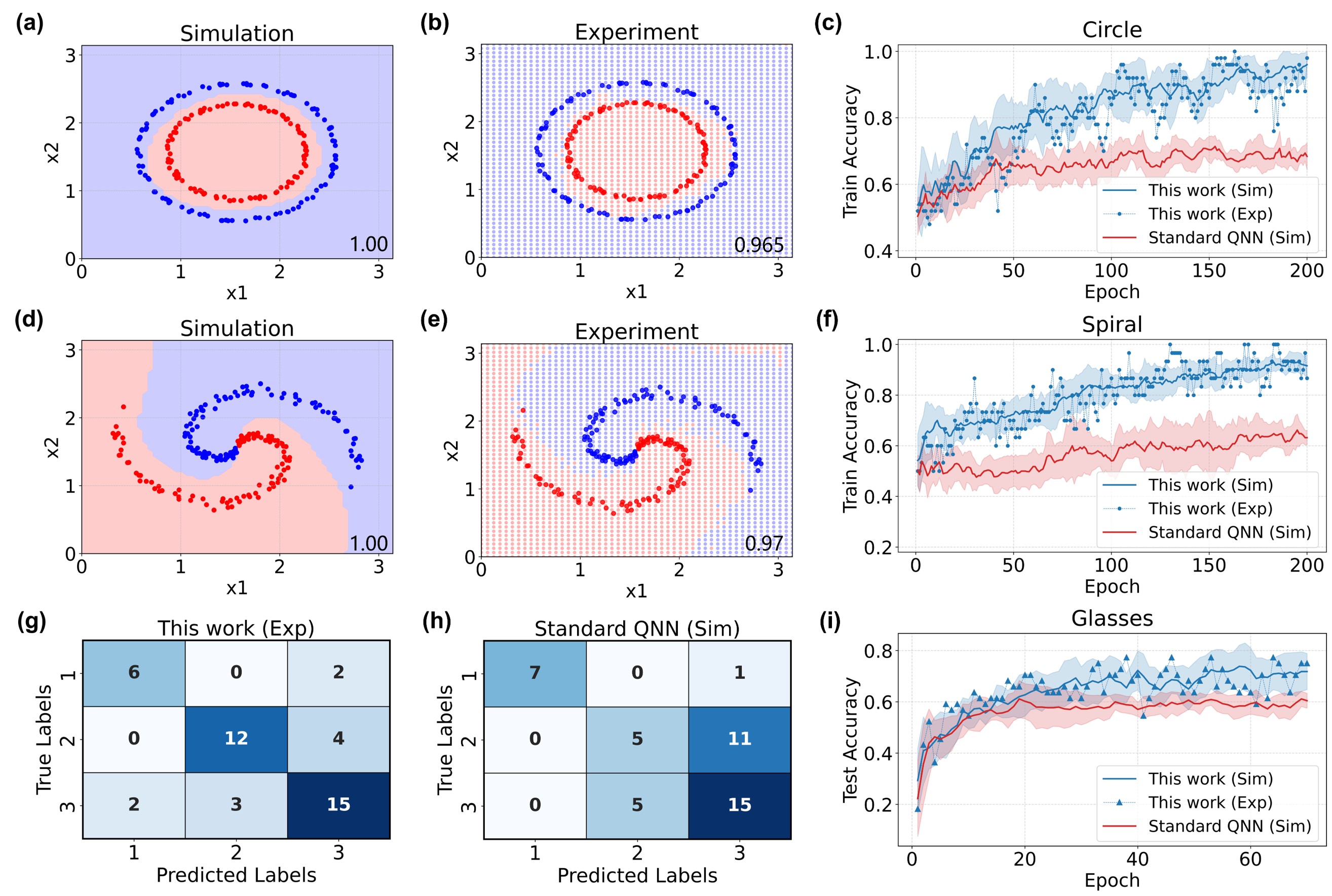}
	\caption{\textbf{Implementation of classification for nonlinear datasets.} (a) Simulated and (b) experimental decision boundaries for the Circle dataset with the nonlinear ADE-QNN. (c) Training accuracy versus epochs, where the blue line and dots represent the simulated and experimental results of the ADE-QNN model, while the red line shows the simulated results for a standard QNN.
		(d)–(f) The corresponding results for the Spiral dataset. (g) Confusion matrix for the Glass dataset obtained from the chip-implemented nonlinear ADE-QNN. Each column represents the instances of a predicted label, and each row represents the instances of the true label. The diagonal elements indicate the number of correctly predicted instances. The chip achieves a testing accuracy of 0.75.
		(h) Simulated confusion matrix of a standard QNN, which achieves a test accuracy of 0.614. (i) Test accuracy versus training epochs for different models. All lines show the mean values of 10 simulated runs with different initial parameters, and the shaded area represents the standard deviation.
	}\label{fig3}
\end{figure*}

To demonstrate the performance of our ADE-QNN chip and validate the design for realizing the quantum circuits, we evaluate several key metrics. First, we characterize the quantum sources, including the photon separation characteristics and interference properties of the bidirectionally pumped quantum light sources, which ensure high generation efficiency and photon indistinguishability. By adjusting the relative phase of the dual pumps inside the sources, we identify the condition that maximizes photon separability. The calculated mean split ratio reaches as high as 0.978(8), as indicated by the brown bars in Figure 2(d). The indistinguishability of each source is evaluated via on-chip Hong-Ou-Mandel (HOM) interference and quantified by the visibility, defined as $V = (CC_{max}-CC_{min})/(CC_{max}+CC_{min})$, where $CC_{max}$ and $CC_{min}$ represent the maximum and minimum coincidence rates, respectively. The green bars in Figure 2(d) show the corresponding HOM visibilities of the four sources, with an average visibility of 0.928(6). More detailed analysis of chip configurations and setups for the characterizations is shown in Supplementary Notes 5. Furthermore, we characterize the quantum interference between different sources to evaluate their mutual indistinguishability. As summarized in Figure 2(e), the measured visibilities for the six source pairs exhibit an average quantum visibility of 0.964(4), demonstrating high indistinguishability across all four sources. Finally, the reconfigurability of the linear network is evaluated by generating and implementing the function of a multiqubit CCCX gate. The CCCX gate can be obtained by setting all unitary operations in the MCRY module to a factor of 1/3. The circuit thus results in the transformation of $|0011\rangle$ to $|0111\rangle$ and vice versa \cite{ralph2002linear}. Figure 2 (f) shows the theoretical and measured input-output truth tables. The performance is quantified using the statistical fidelity $F$, defined as $F = (\sum_{i} \sqrt{p_i q_i})^2$, where $p_i$ and $q_i$ denote theoretical and experimental distributions, respectively. The measured fidelity is 0.924(4), confirming high operational accuracy. The above characterization results show high quality of the source and high programmability of the circuit, laying a solid foundation for the demonstration of ADE-QNN.

\vspace{1em}
\noindent \textbf{Classification of nonlinear datasets} 

\begin{figure*}[!t]
	\centering
	\includegraphics[width=1\textwidth]{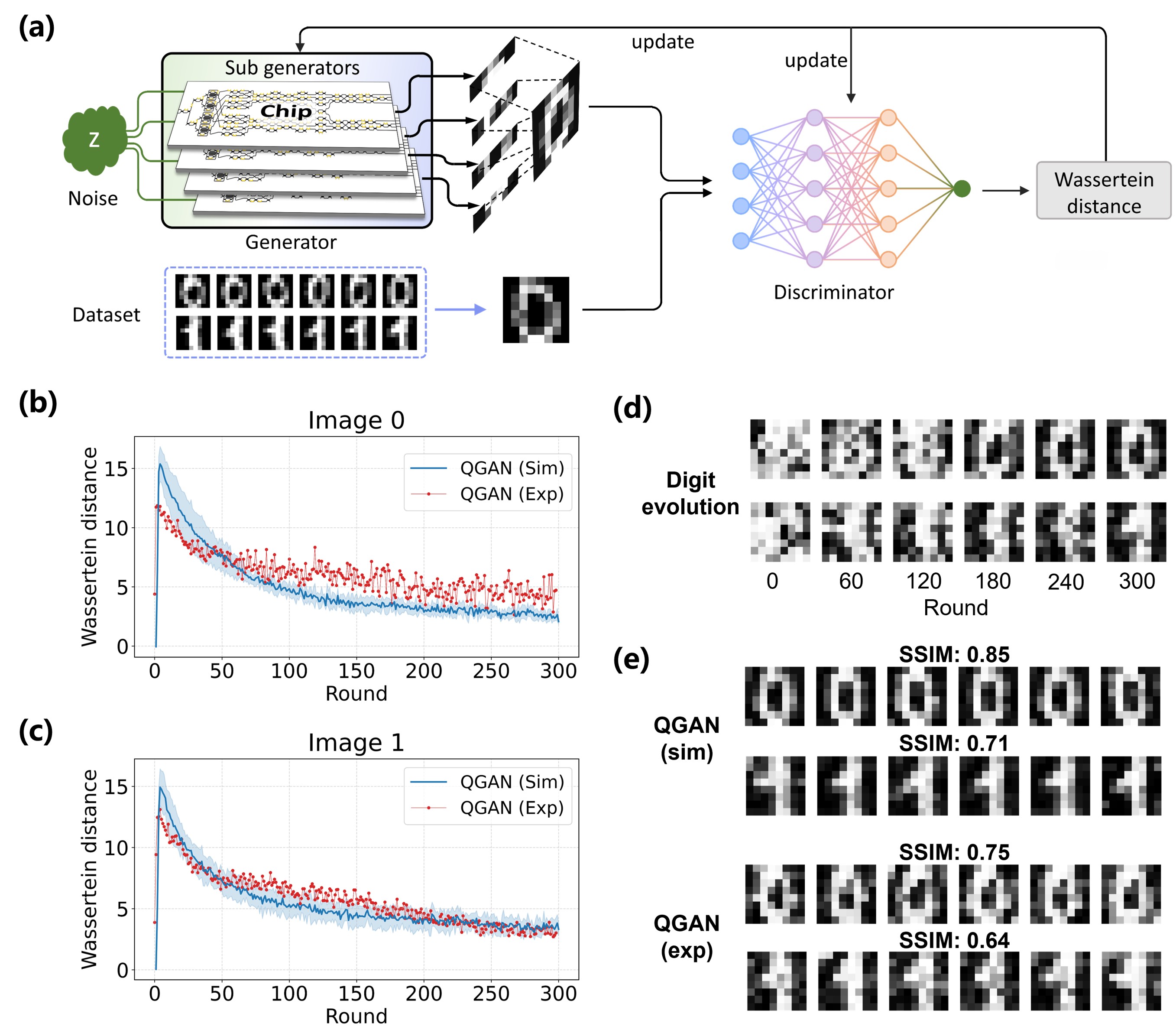}
	\caption{\textbf{Implementation of quantum generative adversarial networks for image generation.} (a) Framework of the patched QGAN for image generation. The generator comprises four sub-generators implemented on the chip, which collectively produce an $8\times8$ image by combining their outputs. (b, c) Wasserstein distance versus the number of training rounds for digit 0 and digit 1. The blue lines and red points represent the simulation and experimental results, respectively. The lines show the mean over 10 runs with different initializations, and the shaded areas indicate the standard deviation. (d) Evolution of the generated digits during the training process. (e) Samples of images generated by the simulated and experimental QGANs.}\label{fig4}
\end{figure*}

To demonstrate the nonlinear capabilities of the ADE-QNN, we train it on three strongly nonlinear datasets: Circle, Spiral, and Glass Identification. The Circle and Spiral datasets each comprise two input features and two output classes. The visualizations in Figure 3(a) and 3(d) confirm that both datasets are linearly nonseparable. Each dataset is split into 50 training samples and 200 test samples. The classification is performed using the two-layer quantum neural network illustrated in Figure 2(b). In detail, the two inputs are first duplicated to match the four-qubit register and then encoded into the rotation angles of the $RY$ gates and $RZ$ gates in the input layer and the first hidden layer, respectively. As mentioned before, this replication of input data into the same quantum layer multiple times enhances nonlinearity, and detailed analysis can be found in Supplementary Note 1 and Ref. \cite{wu2021expressivity, jerbi2023quantum}. The remaining parameterized gates in the network are trained to map inputs onto two output modes ($q_2$ and $q_3$), whose expectation values are measured via photon detection. A one-hot encoding scheme is used to indicate the subspecies labels. In the training process, the mean squared error (MSE) is adopted as the loss function, and the SPSA-AMSGrad algorithm is applied for the gradient evaluation and parameter optimization (see details in Supplementary Note 2). For comparison, we trained a standard linear QNN without data replication and MCRY modules \cite{bharti2022noisy} (see Supplementary Note 6). Evidently, our ADE-QNN exhibits a much higher training accuracy upon convergence (blue solid curves), than the standard QNN (red solid curves) in Figure 3(c) for Circle and 3(f) for Spiral, indicating a substantial performance enhancement. The experimental convergence of accuracy versus epochs for the ADE-QNN, depicted as blue solid points in Figure 3(c) and 3(f), aligns closely with the simulated blue solid curves. Upon the completion of the training, our ADE-QNN demonstrates high classification accuracies of 0.967 and 0.97 on the Circle and Spiral datasets, drastically outperforming the linear network (0.714 and 0.662). The resulting decision boundaries visualized in Figure 3(b) and 3(e), where dots indicate ground truth and background color represents predictions, visually confirm this substantial enhancement, underscoring the critical role of our architecture's nonlinear capabilities.

A more complicated task, the Glass Identification, is used to further validate the nonlinearity of our ADE-QNN chip. The original dataset contains 214 glass samples characterized by 9 attributes and classified into 6 types. Due to the chip’s size limitation, we deliberately select three typical classes of glass with 175 instances and compress the input features into four dimensions (see Supplementary Note 6 for details). The selected dataset is split into the training set and testing set according to a ratio of 75:25. The training spans 70 epochs with a batch size of 32, resulting in 5 batches per epoch. The simulation and experimental training processes are presented in Figure 3(i) as solid lines and points, respectively, and show a clear accuracy improvement when the nonlinear modules are activated. The final experimental accuracy on 44 test samples is 0.75 for the ADE-QNN (Figure 3(g)), far exceeding 0.614 obtained in the simulation with the traditional QNN (Figure 3(h)). These results demonstrate that our ADE-QNN provides significant nonlinearity for handling complex classification tasks.

\vspace{1em}
\noindent \textbf{Image generation using quantum generative adversarial networks (QGANs)} 

A more complicated task, generating images, is further achieved by a QGAN with our ADE-QNN chip. In this task, we build a QGAN and train it on the digit dataset (MNIST dataset) to generate similar images with $8\times8$ pixels \cite{dua2017uci}. Our chip supports four computational qubits, yielding pixel vector of input images with $2^4=16$ via the probabilities $\{p_j\}$ of the computational basis states. To generate larger $8\times8$ (64-pixel) images, we encode the data across four chips and patch the resulting outputs together. This demonstrates that generative tasks requiring higher-dimensional outputs can be solved by using multiple chips in parallel or reusing a single chip iteratively \cite{huang2021experimental, sedrakyan2024photonic}. Further details on this patched method are provided in Supplementary Notes 7. Following the routine of classical generative adversarial networks (GANs) \cite{goodfellow2014generative}, our QGAN model based on the ADE-QNN chip is illustrated in Figure 4(a), in which a two-layer minimax game between a quantum generator $G$ and a classical discriminator $D$ is exploited. The generator $G$ learns the patterns from sampled images and generates fake images $G(\vec{Z})$ from a noise input, and the discriminator $D$ then evaluates the similarity between the distributions of real and generated fake images. To ensure stable training and mitigate common GAN failures such as mode collapse, we employ the Wasserstein distance with gradient penalty \cite{gulrajani2017improved}, which alters the training dynamics to prevent the discriminator from overpowering the generator. In our experiment, a two-dimensional noise vector $\vec{Z}$, sampled uniformly from [0, 1], is encoded into the rotation angles of the gates for qubits $q_0$ and $q_2$. The training model, implemented in PyTorch, employs the RMSProp optimizer for both the quantum generator and the classical discriminator. Training spans 300 iterations with a batch size of 4. The classical discriminator utilizes a multilayer perceptron neural network architecture comprising two hidden layers, while the quantum generator is implemented using our patched ADE-QNN chips. 

Figure 4(b) and 4(c) show the training dynamics of the discriminator $D$ for generating digits "0" and "1". The decreasing Wasserstein distance, for both simulation (blue lines) and experiment (red points), indicates successful QGAN training. The experimental results show marginally higher distances, primarily due to the statistical noise inherent in estimating the image distribution with a finite number of sampling shots. Crucially, the visual progression from noise to recognizable digits in Figure 4(d) demonstrates the practical effectiveness of our ADE-QNN for generative modeling on a quantum photonic chip. We evaluated the quality of the final generated images (Figure 4(e)) using the Structural Similarity Index Measure (SSIM) \cite{wang2004image}. The achieved SSIM scores, exceeding 0.71 (simulation) and 0.64 (experiment), confirm high-fidelity generation. This performance is particularly notable given the inherent complexity and high parameter cost of generative tasks \cite{goodfellow2020generative}. Our ADE-QNN achieves this competitive result with exceptional parameter efficiency, delivering superior image quality to the boson sampling scheme in Ref. \cite{sedrakyan2024photonic} and matching the benchmark QNN in Ref. \cite{huang2021experimental}, despite using only two entanglement layers and four qubits compared to the latter's four layers and five qubits, as provided in Table 1. These results collectively affirm the high expressive power and capability of our ADE-QNN in executing complex generative tasks.

\vspace{1em}
\noindent \textbf{Gibbs state generation with quantum generative diffusion model (QGDM)} 

Beyond classical problems where our ADE-QNN demonstrates enhancement, numerous challenges in quantum science, ranging from basic state generation \cite{horodecki2009quantum} to complex many-body physics simulations \cite{aspuru2005simulated}, lie beyond the reach of classical techniques, necessitating a native quantum approach. Our ADE-QNN chip is also uniquely suited for such tasks. As a demonstration of its capabilities, we employ it to implement a QGDM\cite{zhang2024generative, kwun2025mixed, chen2024quantum} for the preparation of Gibbs states, a class of quantum states vital to studies in quantum thermodynamics and many-body systems \cite{lostaglio2015description}. To the best of knowledge, this is the first experimental QGDM running on a real quantum device.

\begin{figure*}[!t]
	\centering
	\includegraphics[width=1\textwidth]{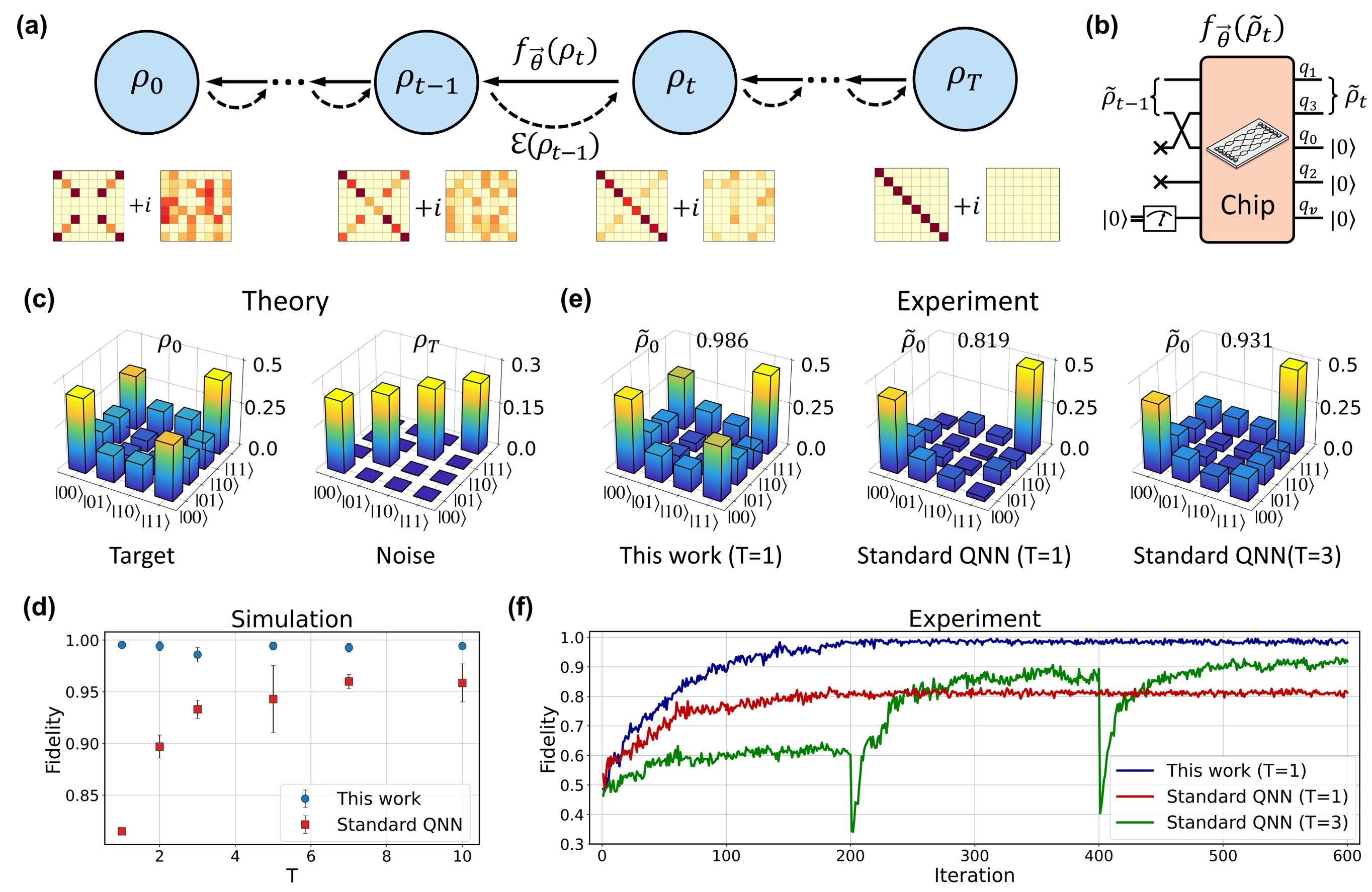}
	\caption{\textbf{Implemention of QGDM for Gibbs state generation.} (a) Schematic of the QGDM. It consists of a forward diffusion process $\mathcal{E}(\rho_{t-1})$ and a backward denoising process $f_{\vec{\theta}_t}(\tilde{\rho}_{t})$, with each process comprising $T$ distinct time steps. (b) Implementation of $f_{\vec{\theta}_t}(\tilde{\rho}_{t})$ on our chip. At each time step, $\tilde{\rho}_{t}$ is loaded into the state of $q_3q_1$ and transformed into $\tilde{\rho}_{t-1}$ on $q_1q_0$. The qubits $q_3q_2$ remain unmeasured, while the virtual qubit ($q_{\bm{v}}$) is post-selected for the outcome $|0\rangle$. (c) Density matrices of the target 2-qubit Gibbs state (left) and a maximally mixed noise state (right).
		(d) Simulated fidelities for different values of $T$. The blue and red points represent the results from our ADE-QNN and the standard QNN, respectively. (e) Experimentally reconstructed density matrices (from left to right): our ADE-QNN at $T=1$, and the standard QNN at $T=1$ and $T=3$, with corresponding fidelities of 0.986, 0.819, and 0.931 for reconstructed Gibbs state. (f) Experimental fidelity curves for the three configurations.}\label{fig5}
\end{figure*}

The QGDM, shown in Figure 5(a), includes a forward diffusion process $\mathcal{E}(\rho_{t-1})$ executed on a classical computer and a backward denoising process $f(\vec{\theta})(\tilde{\rho}_t)$ (Figure 5(b)) implemented on our chip. The forward process applies $T$ iterative steps of noise injection via a non-unitary depolarizing channel, transforming the input state $\rho_0$ into a maximally noisy state $\rho_T$, the completely mixed state $I/d$ for a $d$-dimensional system. The backward process aims to reconstruct $\tilde{\rho}_0$ from $\rho_T$ by training $T$ distinct denoising functions $f(\vec{\theta})(\tilde{\rho}_t)$, which are executed by applying the quantum chip sequentially $T$ times. 
At denoising step $t$, the chip receives $\tilde{\rho}_t$ via qubits $q_1q_3$ and output $\tilde{\rho}_{t-1}$ on $q_0q_1$ via quantum state tomography with maximum likelihood estimation \cite{lee2024quantum}, which is subsequently fed back into the $q_1q_3$ ports of the chip for the next denoising step. Note that, for this application, the input layer is realized by the initial operations $U{s_1}$, $U_{s_2}$, and $U_{s_3}$ on $q_1q_3$ (Figure 2(a) and 2(b)), which can generate arbitrary two qubit states \cite{ma2024quantum}, enabling efficient quantum state loading. Since the forward noise injection is non-unitary, our implementation requires non-unitary operations. We achieve this by tracing out the entangled subsystem $q_2q_3$ and performing post-selection on the virtual qubits $q_{\bm{v}}$. The training objective at each step is to maximize the fidelity between the denoised state $\tilde{\rho}_t$ and the target state $\rho_t$ from the forward process. After training all $T$ steps, the final output $\tilde{\rho}_0$ approximates the original state $\rho_0$ (see Supplementary Note 8 for details). The approximation error typically decreases as $T$ increases \cite{zhang2024generative, kwun2025mixed, chen2024quantum}. In our experiment, the target state is a two-qubit Gibbs state, and the added noise is the maximally mixed state; their density matrices are shown in Figure 5(c). We trained the QGDM using two networks for state reconstruction from noise: namely, the ADE-QNN described above and a standard QNN (as used in classification tasks) that relies solely on partial trace operations to introduce non-unitary evolution \cite{chen2024quantum}.
The simulated state fidelities after training, as a function of the number of diffusion steps $T$, are presented in Figure 5(d). The results show that our ADE-QNN architecture (this work, blue points) achieves a near-unity fidelity at $T=1$, while the standard QNN demonstrate lower fidelities 0.96 even at $T=10$. This result confirms the general capability and superior expressivity of our architecture for implementing non-unitary quantum state evolution.

\begin{table*}[!t]
	\centering
	\footnotesize
	
	\begin{threeparttable}
		
		\caption{Comparison of quantum machine learning models.}
		
		\begin{tabularx}{\textwidth}{
				>{\centering\arraybackslash}p{1cm}  
				>{\centering\arraybackslash}p{1.5cm}  
				>{\centering\arraybackslash}p{1.4cm}  
				>{\centering\arraybackslash}p{1.1cm}  
				>{\centering\arraybackslash}p{3.2cm}  
				>{\centering\arraybackslash}p{1.6cm}  
				>{\centering\arraybackslash}p{1.8cm}  
				>{\centering\arraybackslash}p{1.6cm}  
				>{\centering\arraybackslash}p{1.7cm}  
				>{\centering\arraybackslash}p{1.6cm}  
			}
			
			\toprule
			&&&&&&& \multicolumn{3}{c}{Result} \\
			\cmidrule(lr){8-10}
			
			Ref. & Model & Platform & Encoding & Nonlinearity & Non-unitary & Resources & Task & Dataset & Performance \\
			\midrule
			
			\cite{hoch2025quantum} & Kernel & P & BS & Measurement-based feedback & -- & 3-photon & C & 2D Moon & 0.90 \\[1.8em]
			\cite{yin2025experimental} & Kernel & P & BS & Data encoding and measurement & -- & 2-photon & C & Custom & 0.80 \\[1.8em]
			\cite{maring2024versatile} & QNN & P & BS & Measurement & -- & 3-photon & C & Iris & 0.95 \\[0.8em]
			\cite{sedrakyan2024photonic} & QNN & P & BS & Measurement & -- & 3-photon & QGAN & MNIST & -- \\[0.8em]
			\cite{huang2021experimental} & QNN & S & Qubit & Ancillary qubit and measurement & -- & 5-qubit & QGAN & MNIST & -- \\[1.8em]
			\cite{monbroussou2025photonic} & PCQNN & P & BS & Measurement-based feedback & -- & 4-photon & C & BAS & 0.909 \\[1.8em]
			
			\cmidrule{8-10}
			\multicolumn{2}{c}{\multirow{3}{*}{\vspace{-2em}\shortstack{ADE-QNN \\ (This work)}}} & \multirow{3}{*}{\vspace{-2em}P} & \multirow{3}{*}{\vspace{-2em}Qubit} & \multirow{3}{*}{\vspace{-2em}\shortstack{Ancillary high- \\ dimensional space}
			} & \multirow{3}{*}{\vspace{-2em}\shortstack{MCRY}
			} &  \multirow{3}{*}{ \vspace{-2em} \shortstack{2-photon \\ (4-qubit)}} & C & 2D Spiral & 0.97 \\[0.1em]
			\cmidrule{8-10}
			\multicolumn{2}{c}{} & & & & & & QGAN & 8$\times$8 MNIST & 0.75 \\[0.5em]
			\cmidrule{8-10}
			\multicolumn{2}{c}{} & & & & & & QGDM &  Gibbs state & 0.986 \\
			
			\bottomrule
			
		\end{tabularx}
		
		\begin{tablenotes}
			\item P:photonics; S:superconducting; BS:bonson sampling; C:classification. Note that the metric for the classification task denotes the test set accuracy, while for the QGAN task, it refers to the Structural Similarity Index (SSIM), and for the QGDM task, it indicates the fidelity.
		\end{tablenotes}
		
	\end{threeparttable}
\end{table*}

Furthermore, the experimental training results for our ADE-QNN and the standard QNN are shown in Figure 5(e) and 5(f), respectively. 
At the case of $T=1$, namely directly learning the target state from noise, our ADE-QNN (blue curve in Figure 5(f)) achieves a final fidelity of 0.986, whereas the standard QNN (red curve) reaches only 0.819 for the reconstructed Gibbs state; the corresponding two density matrices are displayed in Figure 5(e). To further demonstrate the QGDM's capability to enhance state fidelity, we extended the diffusion process of the standard QNN to $T=3$ steps (green curve, Figure 5(f)). At the start of each new denoising step, the fidelity drops sharply due to random parameter initialization before increasing to surpass the final fidelity of the previous step. The three steps achieve convergent fidelities of 0.644, 0.907, and 0.931, respectively. The final generated density matrix (right panel of Figure 5(e)) has a fidelity of 0.931, which is still substantially lower than the result achieved with our ADE-QNN architecture. Detailed experimental parameters are provided in Supplementary Note 8 and Supplementary Table 1. Thus, our ADE-QNN enables highly efficient quantum state generation, providing a powerful architecture for implementing non-unitary quantum state evolution in complex quantum systems.

\vspace{1em}
\noindent \textbf{Discussion}

In this work, we propose and experimentally demonstrate a new approach to achieve highly efficient ADE-QNN, where the adding of ancillary high-dimensional computation space overcomes the critical limitation of qubit resource consumption, by leveraging input replication and mode expansion, all within a linear quantum photonic circuit. Unlike previous methods reliant on  probabilistic multi-qubit unitary gates for a linear operation and  ancillary qubits or measurement-based feedback for a nonlinear evolution, both of  which incur low success rates, high qubit overhead and measurement loss, our approach allows the construction of deep QNNs with effective linear and nonlinear operations without resorting to additional physical qubits or requiring layer-by-layer measurements. Specifically, we eliminate the reliance of linear layers on probabilistic circuits by introducing non-unitary operations, while achieving nonlinearity through resource-efficient data re-encoding. Although our non-unitary gates still incur inherent photon loss, these losses are significantly lower than the failure rates associated with probabilistic gate circuits, thereby substantially improving system cascadability. Simulation results (detailed in Supplementary Note 9) further confirm the superior performance of our design in deep cascading and high-accuracy classification.

This ADE-QNN scheme is further experimentally validated on a programmable linear silicon-photonic quantum chip, which integrates four high-fidelity entanglement sources with a programmable high-dimensional interference network. This hardware successfully implemented a two-hidden-layer ADE-QNN, and its design is primed for future expansion. Combining the theoretical scheme with the experimental hardware platform, our chip resolves various machine learning tasks, including classification of nonlinear two-label and three-label datasets, image generation via a Patch QGAN, and quantum Gibbs state learning using a QGDM. These results confirm that dimension expansion and input data reuploading unlock superior expressivity in our ADE-QNN, empowering a fundamentally linear quantum optical system to execute nonlinear tasks. In addition, a comparison of our chip-implemented ADE-QNN with other representative works, including kernel methods and photonic conventional quantum neural network (PCQNN) schemes, is provided in Table 1. Our model demonstrates significant advantages in both resource consumption and classification accuracy, outperforming other state-of-the-art approaches. Furthermore, these demonstrations highlight the scheme's potential for practical applications in image generation and highly efficient quantum state preparation. Looking forward, the proposed MCRY unit can offers a path toward deeper networks via controllable gain compensation to offset photon loss, while chip-multiplexing allows the platform to scale for large-scale data processing and multi-time-step evolution.

In conclusion, our work provides a foundational framework and a versatile integrated photonic platform that opens new avenues for large-scale, multilayer QNNs. We envision that our approach and the designed chip will help solve problems beyond the reach of classical methods and pave the way for the next generation of quantum deep learning systems.

\section*{Methods}

\noindent \textbf{State evolution on the MCRY module}
 
The input state before the MCRY module is a 4-qubit state on the computational qubits, given by 
\begin{equation}
	|\Phi\rangle = \sum\limits_{x_3,x_2\in{0,1}}\sum\limits_{x_1,x_0\in{0,1}}\alpha_{\bm{x}}|x_3x_2\rangle_{32}|x_1x_0\rangle_{10},
\end{equation}

\noindent where $\bm{x}=x_3x_2x_1x_0$ denotes the binary representation of the computational basis, and $\alpha_{\bm{x}}$ are complex coefficients. To simplify the analysis, we first map the two-qubit states onto a four-dimensional qudit state in each single photon as follows:
\begin{equation}
	\left\{
	\begin{aligned}
		&|00\rangle_{10(32)}\to|0\rangle_{p_1(p_2)}\\
		&|01\rangle_{10(32)}\to|1\rangle_{p_1(p_2)}\\
		&|10\rangle_{10(32)}\to|2\rangle_{p_1(p_2)}\\
		&|11\rangle_{10(32)}\to|3\rangle_{p_1(p_2)},\\
	\end{aligned}
	\right.
\end{equation}

\noindent where $p_1$ and $p_2$ denote the two photons, and $10$ and $32$ are the indices of the computational qubits they represent. In the MCRY module, the Hilbert space is expanded by a factor of four through the extension of specific optical paths for two photons. This process is mathematically equivalent to the introduction of two additional qubits (the virtual qubits) into the system. By labeling the original paths of two photons as $|0\rangle_{v_1}$ and $|0\rangle_{v_2}$ and the expanded waveguides as $|1\rangle_{v_1}$ and $|1\rangle_{v_2}$, the input state can be rewritten as
\begin{equation}
	\begin{aligned}
		|\Phi\rangle^{in} &= \sum\limits_{i=0}^{3}\sum\limits_{j=0}^{3}\alpha_{ij}|i\rangle_{p_2}|j\rangle_{p_1}|0\rangle_{v_2}|0\rangle_{v_1}\\
		&=|\phi\rangle_0^{in}+|\phi\rangle_1^{in}+|\phi\rangle_2^{in}.
	\end{aligned}
\end{equation}

Here, $|\phi\rangle_0^{in}$, $|\phi\rangle_1^{in}$, and $|\phi\rangle_2^{in}$ denote three distinct types of states, corresponding to 0, 1, and 2 photons passing through the MZI with $\theta_z$, respectively. These states undergo different transformations in the MCRY module. Their detailed expansions in the computational basis are given by
\begin{equation}
	\left\{
	\begin{aligned}
		&|\phi\rangle_0^{in} = \sum\limits_{i=1}^{3}\sum\limits_{j=0}^{2}\alpha_{ij}|i\rangle_{p_2}|j\rangle_{p_1} |0\rangle_{v_2}|0\rangle_{v_1}\\
		&|\phi\rangle_1^{in} = \left(|0\rangle_{p_2}\sum\limits_{j=0}^{2}\alpha_{0j}|j\rangle_{p_1} +\sum\limits_{i=1}^{3}\alpha_{i3}|3\rangle_{p_2}|0\rangle_{p_1}\right)|0\rangle_{v_2}|0\rangle_{v_1}\\
		&|\phi\rangle_2^{in} = \alpha_{03}|0\rangle_{p_2}|3\rangle_{p_1}|0\rangle_{v_2}|0\rangle_{v_1}
	\end{aligned}.
	\right.
\end{equation}

We first derive the state evolution of $|\phi\rangle_{0}^{\rm in}$. Considering the subspace associated with the MZI having angle $\theta_0^{p_1}$, the MZI performs an RY-like operation, transforming the input state $|00\rangle_{10}|0\rangle_{v_1}$ into $|00\rangle_{10}\otimes R_y(\theta_0^{p_1})|0\rangle_{v_1} = |00\rangle_{10}\left[\sin(\theta_0^{p_1}/2)|0\rangle_{v_1}+\cos(\theta_0^{p_1}/2)|1\rangle_{v_1}\right]$, where the component with $|1\rangle_v^{p_1}$ is discarded and cannot be measured. Here, the RY gate in our framework is defined as
\begin{equation}
	R_y(\theta)=
	\begin{pmatrix}
		\sin(\theta/2) & \cos(\theta/2) \\
		\cos(\theta/2) &  -\sin(\theta/2)
	\end{pmatrix},
\end{equation}

\noindent which expresses the matrix transformation of a MZI.

Extending this procedure to the sequence $\theta_1^{p_1}, \theta_2^{p_1}$ and $\theta_1^{p_2}-\theta_3^{p_2}$, the output state of $|\phi\rangle_{0}^{\rm in}$ after the MCRY module is therefore
\begin{equation}
	\begin{aligned}
		|\phi\rangle_{0}^{out} &= \sum\limits_{i=1}^{3}\sum\limits_{j=0}^{2}\alpha_{ij}|i\rangle_{p_2}|j\rangle_{p_1} \otimes R_y(\theta_{i}^{p_2})|0\rangle_{v_2}\otimes R_y(\theta_{j}^{p_1})|0\rangle_{v_1}.
	\end{aligned}
\end{equation}

This operation can be effectively modeled by the six controlled RY gates shown in Fig. 2(c) (three on the left and three on the right).

Next, we analyze the state evolution of $|\phi\rangle_{1}^{\rm in}$, in which one photon enters the MZI with $\theta_z$. Due to the two-photon coincidence condition, only the transmission outcomes ($|0\rangle_{v_1}$ and $|0\rangle_{v_2}$) are recorded, while the crossing outcomes ($|1\rangle_{v_1}$ and $|1\rangle_{v_2}$) are discarded. Consequently, the output state becomes
\begin{equation}
	\begin{aligned}
		|\phi\rangle_{1}^{out} &=|0\rangle_{p_2}\sum\limits_{j=0}^{2}\alpha_{0j}|j\rangle_{p_1}\otimes R_y(\theta_{z})|0\rangle_{v_2}\otimes R_y(\theta_{j}^{p_1})|0\rangle_{v_1}\\ &\quad+\sum\limits_{i=1}^{3}\alpha_{i3}|3\rangle_{p_2}|0\rangle_{p_1}\otimes R_y(\theta_{i}^{p_2})|0\rangle_{v_2}\otimes R_y(\theta_{z})|0\rangle_{v_1}.
	\end{aligned}
\end{equation}

Finally, for the state $|\phi\rangle_{2}^{\rm in}$, we consider $\theta_z \in \{\pi, 2\arcsin(\sqrt{1/3})\}$, which are the two phase values used in our QNN experiments. If $\theta_z = \pi$, both photons transmit through the MZI without changing the state, yielding $|\phi\rangle_{2}^{out} = \alpha_{03}|0\rangle_{p_2}|3\rangle_{p_1}\otimes R_y(\pi)|0\rangle_{v_2}\otimes R_y(\pi)|0\rangle_{v_1}$. If $\theta_z = 2\arcsin(\sqrt{1/3})$, two-photon interference occurs with a coincidence probability of $1/9$ and introduces a minus phase \cite{ralph2002linear}, resulting in $|\phi\rangle_{2}^{out} =\alpha_{03}|0\rangle_{p_2}|3\rangle_{p_1}\otimes R_y(\pi)|0\rangle_{v_2}\otimes R_y(2\arcsin(1/3)+2\pi)|0\rangle_{v_1}$. By defining a function
\begin{equation}
	f(\theta_z) = 
	\begin{cases}
		\pi & \text{if } \theta_z = \pi \\
		2\arcsin1/3+2\pi & \text{if } \theta_z = 2\arctan1/\sqrt{3}.
	\end{cases}
\end{equation}

We can then combine the output state of $|\phi\rangle_{2}^{\rm in}$ as
\begin{equation}
	\begin{aligned}
		|\phi\rangle_{2}^{out} &=\alpha_{03}|0\rangle_{p_2}|3\rangle_{p_1}\otimes R_y(f(\theta_z))|0\rangle_{v_2}\otimes R_y(\pi)|0\rangle_{v_1}.
	\end{aligned}
\end{equation}

The transformation in Equations (9) and (11) is together modeled by the middle part of Fig. 2(c) within the box. Without loss of generality, Equations (6)-(11) yield the final output state of the MCRY module as
\begin{equation}
	\begin{aligned}
		|\Phi\rangle^{out} &=|\phi\rangle_{0}^{out}+|\phi\rangle_{1}^{out}+|\phi\rangle_{2}^{out}\\
		&=\sum\limits_{i=0}^{3}\sum\limits_{j=0}^{3}\alpha_{ij}|i\rangle_{p_2}|j\rangle_{p_1}\otimes R_y(\theta_{ij}^{p_2})|0\rangle_{v_2}\otimes R_y(\theta_{ij}^{p_1})|0\rangle_{v_1}\\
		&=MCRY|\Phi\rangle^{in}.
	\end{aligned}
\end{equation}

Therefore, the MCRY module performs multiple controlled RY gates. Note that, during the measurement stage, only photons transmitting through the MCRY module—namely, those in the $|0\rangle_{v_2}|0\rangle_{v_1}$ state—can be detected, resulting in the effective part of the state being
\begin{equation}
	\begin{aligned}
		|\Phi\rangle^{final}
		=&\frac{1}{N(\bm{\theta})}
		\sum_{i=0}^{3}\sum_{j=0}^{3}
		\alpha_{ij}
		|i\rangle_{p_2}|j\rangle_{p_1}
		\\
		&\otimes
		\sin\!\left(\theta_{ij}^{p_2}/2\right)|0\rangle_{v_2}
		\otimes
		\sin\!\left(\theta_{ij}^{p_1}/2\right)|0\rangle_{v_1}
		\\
		=&
		\left(
		\mathcal{I}\otimes |0\rangle_v\langle0|_v
		\right)
		\mathrm{MCRY}
		|\Phi\rangle^{\mathrm{in}}.
	\end{aligned}
\end{equation}

\noindent where $N(\bm{\theta})$ is a normalization coefficient. It is clear that the state in Equation (13) is a partially measured state. Furthermore, if all the phase values ${\theta_{0}^{p_1}-\theta_{2}^{p_1}, \theta_{1}^{p_2}-\theta_{3}^{p_2}, \theta_z}$ are set to $2\arcsin(1/\sqrt{3})$, the state undergoes a $CCCZ$ gate, yielding
\begin{equation}
	\begin{aligned}
		|\Phi\rangle^{\mathrm{CCCZ}}
		=&\frac{1}{N(\bm{\theta})}
		\Bigg[
		\frac{1}{3}
		\sum_{\substack{i,j=0 \\ ij\neq03}}^{3}
		\alpha_{ij}
		|i\rangle_{p_2}|j\rangle_{p_1}
		\\
		&\qquad
		-\frac{\alpha_{03}}{3}
		|0\rangle_{p_2}|3\rangle_{p_1}
		\Bigg]
		|0\rangle_{v_2}|0\rangle_{v_1}
		\\
		=&
		\mathrm{CCCZ}
		|\Phi\rangle^{\mathrm{in}}.
	\end{aligned}
\end{equation}

\noindent which can be easily transformed to a CCCX gate by inserting two Hadamard gates using the relation $HZH = X$.

\vspace{1em}
\noindent \textbf{Experimental setup}

Two continuous-wave lasers (Santec-510 and Santec-570), operating at wavelengths of 1547.62 nm and 1554.28 nm, respectively, are employed as classical pump sources. Each laser output is first passed through a polarization controller and then combined into a single path using a 50:50 fiber beam splitter. The combined light is subsequently amplified to 19 dBm by an erbium-doped fiber amplifier, followed by an optical isolator to suppress back-reflected spurious light from the chip. To further purify the spectra, two cascaded dense wavelength-division multiplexers are inserted in series. The pump beams are coupled into the chip through grating couplers to excite four bidirectionally pumped quantum light sources \cite{ma2025scheme}, generating photon pairs at 1550.92 nm. A current source is used to drive the phase shifters, which control the behavior of the photons. After passing through the chip, single photons are coupled into an array of fiber-coupled superconducting nanowire single photon detectors (SNSPDs, Photec), which exhibit an average detection efficiency of 90\%. Residual pump photons are suppressed using DWDM filters, while photon coincidence counts are registered by a multichannel TimeTagger. A classical computer is used both to process TimeTagger data and to program the chip for quantum state control and execution of QNN algorithms.

\bibliography{sn-bibliography.bib}

\vspace{1em}
\noindent \textbf{Acknowledgements}

We acknowledge Dr. Dawei Wang for useful discussions.
J. Y. Y. acknowledges the funding support from the  National Natural Science Foundation of China (U22A2082); Jinhua Science and Technology Program (2024-1-021). Y. H. W. acknowledges the funding support from the National Key Research and Development Program of China (2024YFE0211800); Shaoxing Industrial Technology Research and Development Projects (2024B1009); Ningbo Science and Technology Program (2023Z073); ”Leading Goose” R\&D Program of Zhejiang Province (2024C01112); ”Vanguard” R\&D Program of Zhejiang Province (2025C01043, 2025C01009). H. H. Z. acknowledges the funding support from the National Natural Science Foundation of China (No. 62505271); Zhejiang Provincial Natural Science Foundation of China (LMS26F050005); Leading Innovative and Entrepreneur Team Introduction Program of Zhejiang (2025R01004); by the Open Fund of the State Key Laboratory of Integrated Optoelectronics. M. G. acknowledges the funding support from the National Research Foundation of Singapore through the NRF Investigatorship Program (Award No. NRF-NRFI09-0010). 

\vspace{1em}
\noindent \textbf{Author contributions}

H.R.M., H.H.Z., Z.C.Z. and L.A.Y. jointly conceived the idea, H.R.M., H.H.Z., Q.S.L.,B.J.H. and J.Y.Y. performed the numerical simulation and theorectical analysis. H.R.M, H.H.Z., Z.C.Z., Q.S.L., and B.J.H did the experiments. H.R.M., J. G., L.C.K., M.G., J.T., T.C., Y.H.W., and J.Y.Y. were involved in the discussion and data analysis. H.R.M., H.H.Z., J. G., W. L., L.C.K., M.G., J.T., T.C., and J.Y.Y. prepared the manuscript. H.H.Z., L.C.K., T.C., and J.Y.Y. supervised and coordinated all the work. All authors commented on the manuscript.

\end{document}